\documentclass[epj,nopacs]{svjour}
\usepackage{graphics}
\usepackage{amsmath}
\usepackage{amssymb}
\usepackage{amsfonts}
\begin{document}
\title{Supersymmetric gradient flow in 4d ${\cal N}=1$ SQCD}
\author{Daisuke Kadoh\inst{1}%
\thanks{\emph{kadoh@keio.jp}}
\and Naoya Ukita\inst{2}
\thanks{\emph{ukita@ccs.tsukuba.ac.jp}}%
}                     
%
%
\institute{Faculty of Sciences and Engineering, Doshisha University,
            Kyoto 610-0394, Japan
\and 
Center for Computational Sciences, University of Tsukuba, Tsukuba, Ibaraki 305-8577, Japan
}
%
\date{}
%
\abstract{
A supersymmetric gradient flow for four-dimensional ${\cal N}=1$ supersymmetric QCD (SQCD) is proposed.
The flow equation is given in both the superfield and component fields of the Wess-Zumino gauge. 
The superfield flow equation is defined for each of the gauge and matter multiplets individually.
Adding a gauge fixing, the component-field flow equation is defined in the Wess-Zumino gauge in a gauge covariant manner. 
We find that the latter equation is supersymmetric in a sense that the commutator of the flow time derivative and the supersymmetry transformation vanishes up to a gauge transformation.
We also discuss a simplified flow by using the gradient of supersymmetric Yang-Mills (SYM) action instead of using SQCD action
 to define a gauge multiplet flow.
\PACS{
      {PACS-key}{discribing text of that key}   \and
      {PACS-key}{discribing text of that key}
     } 
} 
\maketitle

\section{Introduction} 
\label{sec:introduction}

Gradient flow is a fascinating approach to solving various problems in physics and mathematics.
The Yang-Mills flow \cite{Luscher:2010iy,Luscher:2011bx,Luscher:2013cpa} is widely accepted in lattice gauge theory. 
The most interesting point of gradient flow is that the flowed field correlator is UV-finite at all order of perturbation theory \cite{Luscher:2011bx},
and it leads to many interesting applications 
such as a construction of energy-momentum tensor on a lattice \cite{Suzuki:2013gza,Makino:2014taa}.
So, if the gradient flow approach is extended to supersymmetry (SUSY) cases, further interesting studies could be created. 

For ${\cal N}=1$ supersymmetric Yang-Mills theory (SYM), 
a non-SUSY flow is defined by the Yang-Mills flow for the gauge field \cite{Luscher:2010iy} and a flow 
for the gaugino originally introduced for quark fields in \cite{Luscher:2013cpa}. 
The non-SUSY flow was actually used to create the supercurrent in the perturbation theory \cite{Hieda:2017sqq,Kasai:2018koz} 
and to study SYM numerically \cite{Bergner:2019dim}. 
On the other hand, 
a supersymmetric gradient flow can be introduced by using the gradient of SYM action with respect to the vector superfield \cite{Kikuchi:2014rla}. 
The latter flow can also be given for the component fields of the Wess-Zumino gauge in a gauge covariant and supersymmetric manner \cite{Kadoh:2018qwg}.

In this paper, a SUSY gradient flow is derived in four-dimensional ${\cal N}=1$ SQCD. 
For the gauge multiplet, a flow equation is defined in the similar manner as that of SYM in \cite{Kikuchi:2014rla}, 
while, for the matter multiplet, 
a SUSY flow is defined by imposing superchiral conditions  in the superfield formalism as in the case of the Wess-Zumino model \cite{Kadoh:2019glu}.
Further, the flow can be written in terms of the component fields by taking the Wess-Zumino gauge. 
We find that the flow equation is supersymmetric in a sense that the flow time derivative 
and any supersymmetry transformation commute with each other up to a gauge transformation
as the SYM flow \cite{Kadoh:2018qwg}.

This paper is organized as follows: Sect. 2 introduces ${\cal N}=1$ SQCD in the superfield formalism, Sect. 3 proposes a gradient flow equation of SQCD in the superfield formalism and corresponding flow equations in the component fields of Wess-Zumino gauge, and Sect. 4 is devoted to a summary.

\section{SQCD in the superfield formalism}
Let us consider ${\cal N}=1$ SQCD in Minkowski spacetime.\footnote{
We basically follow the notation used in Ref. \cite{Wess:1992cp}. 
The Greek indices $\mu,\nu$ run from $0$ to $3$.
The metric is given by $\eta_{\mu\nu}={\rm diag}\{-1,+1,+1,+1\}$. 
The four component $(\sigma^\mu)_{\alpha\dot\beta}$ and $(\bar\sigma^\mu)^{\dot\alpha\beta}$ 
are defined as $\sigma^\mu=(-1,\sigma^i)$ and $\bar \sigma=(-1,-\sigma^i)$ where
$\sigma^i$ is the standard Pauli matrix. We use $\sigma^{\mu\nu} \equiv \frac{1}{4}(\sigma^{\mu}\bar\sigma^{\nu}-\sigma^{\nu}\bar\sigma^{\mu})$. 
See  \cite{Wess:1992cp} for other details of the notation and useful 
identities for two-component spinors and $\sigma^\mu$ matrix. 
Also, the Einstein summation convention is used throughout this paper.
}
It is a ${\cal N}=1$ supersymmetric extension of $SU(N_c)$ gauge theory  coupled to matters with $N_c$ colors and $N_f$ flavors. 
The gauge generators $T^a$ $(a=1,\ldots, N_c^2-1)$ are hermitian matrices 
normalized as ${\rm tr}(T^a T^b)=\frac{1}{2}\delta_{ab}$ in the fundamental representation. 
The theory in the Wess-Zumino gauge\footnote{The Wess-Zumino gauge will be given in eqs. (\ref{Ext_gauge_transfomation}) and (\ref{WZ_gauge_fixing}).} consists of a gauge multiplet,
\[(A_\mu^a(x),\lambda^a(x),D^a(x)),\]  
where $A_\mu^a, \lambda_\alpha^a$ and $D^a$ $(a=1,\ldots,N_c^2-1)$ represent a gauge field, a two-component spinor and a real auxiliary field, respectively, 
and also $N_f$ matter multiplets,
\[(\phi_\pm^i(x),q_\pm^i(x),F_\pm^i(x))_f,\ {\rm for}\ f=1,2,\ldots,N_f,\] and  $i=1,\ldots,N_c$ and $\pm$ denoting the fundamental and the antifundamental representations of the gauge group. Here, $(\phi_\pm)^i, (q_\pm)_\alpha^i$ and $(F_\pm)^i$ are a complex scaler field, a two-component spinor and a complex auxiliary field,  respectively. 
For later convenience, 
we introduce Lie algebra valued fields expressed as $A_{\mu}=A_{\mu}^aT^a, \lambda=\lambda^aT^a$ and $D=D^aT^a$.

We focus on the massless $N_f=1$ case as the simplest explanation.  One can straightforwardly extend our result to any case of multi flavors with a non zero mass term. 

The action of ${\cal N}=1$ SQCD in Minkowski spacetime is given by $S_{\rm SQCD}=S_{\rm SYM}+S_{\rm MAT}$: 
\begin{eqnarray}
\label{SYM_action}
&&S_{\rm SYM}=\frac{1}{g^2}\int d^4 x \ 
{\rm tr} \left\{-\frac{1}{2} F_{\mu\nu}^2 
-2 i \bar \lambda \bar\sigma^\mu D_\mu \lambda 
+ D^2 \right\}, \\  
\label{matter_action}
&&S_{\rm MAT}
= \int d^4 x \, \bigg\{
- |D_\mu \phi_+|^2
- |D_\mu \phi_-|^2 + |F_+|^2  + | F_-|^2 \nonumber \\
&&\hspace{1cm} -i \bar q_+\bar\sigma^\mu D_\mu q_+ 
-i q_-\sigma^\mu D_\mu \bar q_- 
+\phi^\dag_+ D\phi_+-\phi_-D\phi^\dag_-  \nonumber \\
&&\hspace{1cm} 
+\sqrt{2} i (
    \phi_+^\dag \lambda q_+ 
+  \phi_- \bar\lambda \bar q_-
 - \bar q_+ \bar\lambda \phi_+ 
-  q_- \lambda \phi^\dag_-
) \bigg\},
\end{eqnarray}
where $S_{\rm SYM}$ is just the action of SYM in the Wess-Zumino gauge and $S_{\rm MAT}$ is the action of the massless matter multiplet interacting with the gauge multiplet.  
In eqs. (\ref{SYM_action}) and (\ref{matter_action}), 
$F_{\mu\nu} =\partial_{\mu}A_{\nu} -\partial_{\nu}A_{\mu} +i[A_{\mu},A_{\nu}]$ and 
$D_\mu$ is a covariant derivative in the representation of acting fields: 
$D_\mu \varphi =\partial_\mu  \varphi + i [A_\mu,  \varphi]$ 
for adjoint fields $\varphi=\lambda$, $D$, $F_{\rho\sigma}$ and 
$D_\mu X_+ =\partial_\mu  X_+ + i A_\mu X_+$ for fundamental fields 
$X_+=\phi_+$, $q_+$, $F_+$ and 
$D_\mu X_- =\partial_\mu  X_- - i X_- A_\mu$ for antifundamental fields $X_-=\phi_-$, $q_-$, $F_-$. 
The action is invariant under an infinitesimal gauge transformation characterized by a Lie algebra valued function $\omega=\omega^aT^a$,
\begin{eqnarray}
\begin{split}
&\delta^g_\omega A_\mu =-D_\mu \omega, \\
&\delta^g_\omega \varphi =i[\omega, \varphi], 
\label{normal_gauge_transf}\\
&\delta^g_\omega X_+ = i \omega X_+, \\
&\delta^g_\omega X_- = X_- (-i\omega) ,
\end{split}
\end{eqnarray}
and a SUSY transformation,
\begin{eqnarray}
&&
 \delta_{\xi} A_{\mu} = i \xi \sigma_\mu \bar  \lambda +  i \bar\xi \bar\sigma_\mu \lambda, 
\nonumber \\
&& \delta_{\xi} \lambda =  \sigma^{\mu\nu} \xi F_{\mu\nu} +i \xi D, 
\nonumber\\
&& \delta_{\xi} D        =  -\xi \sigma^\mu D_\mu \bar  \lambda +  \bar\xi \bar\sigma^\mu D_\mu \lambda,
\nonumber \\
&& \delta_{\xi} \phi_{\pm} =\sqrt{2} \xi q_\pm, 
\label{DF_super}
\\
&& \delta_{\xi} q_{\pm} =\sqrt{2} i \sigma^\mu \bar \xi D_\mu \phi_\pm + \sqrt{2} \xi F_\pm, 
\nonumber \\
&& \delta_{\xi} F_+ = \sqrt{2} i \bar \xi \bar \sigma^\mu D_\mu q_+  +2i \bar \xi \bar \lambda \phi_+,  
\nonumber \\
&& \delta_{\xi} F_- = \sqrt{2} i \bar \xi \bar \sigma^\mu D_\mu q_- -2i \phi_- \bar \xi \bar \lambda,  
\nonumber 
\end{eqnarray}
where $\xi_\alpha, \bar\xi_{\dot\alpha}$ are global anti-commuting parameters.

Let us reconstruct the massless one-flavor SQCD in the superfield formalism.  A superfield ${\cal F}$ is introduced as a function of 
$z=(x^\mu, \theta_\alpha, \bar\theta_{\dot \alpha})$ 
where $\theta_\alpha, \bar \theta_{\dot\alpha}$ are two component Grassmann odd coordinates in addition to the Minkowski coordinates $x^\mu$. It transforms under 
a supersymmetry transformation as 
\begin{eqnarray}
\label{super_transf}
\delta_\xi {\cal F}(z)  = (\xi Q +\bar\xi \bar Q) {\cal F}(z),
\end{eqnarray}
where 
$ Q_{\alpha}$, $\bar Q_{\dot\alpha}$ are differential operators defined by
\begin{eqnarray}
\begin{split}
& 
Q_{\alpha} = 
\frac{\partial}{\partial\theta^{\alpha}} 
-i(\sigma^{\mu})_{\alpha\dot{\alpha}}\bar\theta^{\dot\alpha}\partial_{\mu},
 \\
&
\label{Q_def}
  \bar{Q}_{\dot\alpha} = 
-\frac{\partial}{\partial\bar\theta^{\dot\alpha}} 
+i\theta^{\alpha}(\sigma^{\mu})_{\alpha\dot{\alpha}}\partial_{\mu}.
\end{split}
\end{eqnarray}
For later use, let us introduce other  differential operators $D_{\alpha}$ and $\bar D_{\dot\alpha}$ as
\begin{eqnarray}
\begin{split}
&
D_{\alpha} =
 \frac{\partial}{\partial\theta^{\alpha}}
+i(\sigma^{\mu})_{\alpha\dot{\alpha}}\bar\theta^{\dot\alpha}\partial_{\mu},
\\
&
\label{D_def}
  \bar{D}_{\dot\alpha} =
 -\frac{\partial}{\partial\bar\theta^{\dot\alpha}}
-i\theta^{\alpha}(\sigma^{\mu})_{\alpha\dot{\alpha}}\partial_{\mu}.
\end{split}
\end{eqnarray}
Note that 
$\{Q_\alpha, \bar Q_{\dot \beta}\}=-\{D_\alpha, 
\bar D_{\dot \beta}\}=2i\sigma^\mu_{\alpha\dot\beta}\partial_\mu$ and 
the other anti-commutators vanish.

A chiral superfield $\Lambda$ imposed by a superchiral condition 
$\bar D_{\dot \alpha} \Lambda=0$ may be expanded as
\begin{eqnarray}
\Lambda(y,\theta) = A(y) + \sqrt{2} \theta   \psi(y)
 + \theta \theta  F(y)  
\end{eqnarray}
with $y^\mu = x^\mu + i \theta \sigma^\mu \bar\theta$ and $A, \psi$ and $F$ denote a scaler, a two-component spinor and a complex auxiliary fields.   
Similarly, an anti-chiral superfield $\Lambda^\dag$ is defined by
$D_\alpha \Lambda^\dag=0$.  
Moreover, a vector superfield $V$ that is a Lie algebra valued superfield $V=V^aT^a$ and  satisfies the reality condition $V=V^\dag$ is expanded as 
\begin{eqnarray}
&& V(x,\theta,\bar\theta)= \frac{1}{2}{ C (x) } 
+ i\theta  \chi(x) +\frac{i}{2} \theta \theta  
(M(x) +i N(x)) \nonumber \\
&&\hspace{1.7cm}
-\frac{1}{2}\theta \sigma^\mu \bar\theta  A_\mu(x)
+ i\theta \theta \bar\theta ( \bar\lambda(x) + 
\frac{i}{2} \bar\sigma^\mu \partial_\mu  \chi(x) )
 \nonumber \\
&&\hspace{1.7cm}
+\frac{1}{4} \theta \theta \bar\theta\bar\theta 
( D(x)+ \frac{1}{2}  \square C(x)) + h.c. 
\end{eqnarray}
Here $A_\mu, \lambda, \bar\lambda, D$ are just the component fields of the gauge multiplet in the Wess-Zumino gauge. Extra $C,M,N$ are bosonic fields and $\chi, \bar\chi$ are a two component fermionic fields. 

The SQCD action in the superfield formalism is then written as
\begin{eqnarray}
\label{superfield_sym_action}
&& S_{\rm SYM} = \frac{1}{2g^2}
\int d^4 x \, {\rm tr} \bigg\{\frac{}{}
W^\alpha W_\alpha|_{\theta \theta} + h.c. \bigg\}, \\
\label{superfield_mat_action}
&& S_{\rm MAT} = 
\int d^4 x \, \bigg\{\frac{}{}
Q_+^\dag e^{2V} Q_+ +  Q_-  e^{-2V} Q_-^\dag \bigg\}
\bigg|_{\theta\theta\bar\theta\bar\theta}, 
\end{eqnarray}
where
\begin{eqnarray}
W_\alpha=-\frac{1}{8} \bar D\bar D e^{-2V} D_\alpha e^{2V}
\end{eqnarray}
and $Q_{\pm}$ are chiral superfields that consist of the matter multiplets,  
\begin{eqnarray}
Q_\pm(y,\theta) = \phi_\pm (y) + \sqrt{2} \theta q_\pm (y)  + \theta\theta F_\pm (y). 
\end{eqnarray}
The superfield action has unwanted $C,\chi,M,N$ fields which are not included in 
eqs.~(\ref{SYM_action}) and (\ref{matter_action}) with an enlarged symmetry. 
Eqs.~(\ref{superfield_sym_action}) and (\ref{superfield_mat_action}) are actually invariant under the SUSY transformation of the superfields $V, Q_\pm$ defined in eq.~(\ref{super_transf}),
and an extended gauge transformation generated by a chiral superfield $\Lambda$ as
\begin{eqnarray}
\label{Ext_gauge_transfomation}
\begin{split}
&e^{2V^\prime} =e^{2\Lambda^\dag} e^{2V} e^{2\Lambda}, \\
 \label{egauge}
&Q_+^\prime = e^{-2\Lambda} Q_+, \\
&Q_-^\prime = Q_- e^{2\Lambda}.
\end{split}
\end{eqnarray}
To remove the extra $C,\chi, M,N$ fields in the vector superfield, one can choose the component fields of $\Lambda$ by hand such that 
\begin{equation}
C = \chi = M = N = 0,
\label{WZ_gauge_fixing}
\end{equation}
which is called the  Wess-Zumino gauge. 
With this gauge fixing, 
the superfield action of eqs.~(\ref{superfield_sym_action}) and (\ref{superfield_mat_action}) reproduces the component one of 
eqs.(\ref{SYM_action}) and (\ref{matter_action}) perfectly. 
Then the enlarged symmetry (\ref{super_transf}) and (\ref{Ext_gauge_transfomation}) keeping this gauge 
also reproduces the gauge and SUSY transformations, eqs. (\ref{normal_gauge_transf}) and (\ref{DF_super}).

Just one note that 
although the gradient flow is given by a kind of diffusion equation in Euclidean spacetime, 
we derive it in Minkowski spacetime with the superfield formalism and then expand it in terms of the component fields.  
In order to obtain the gradient flow in Euclidean spacetime, one should performs the Wick 
rotation $t \rightarrow -it$, $A_0 \rightarrow i A_0$ and the replacement of the auxiliary fields $D \rightarrow iD$, $F_\pm \rightarrow i F_\pm$ and
$F^\dag_\pm \rightarrow i F^\dag_\pm$.

\section{Derivation of SQCD flow}

We introduce a fictitious flow time $t$ and $t$-dependent superfields  to define a gradient flow in the superfield formalism.  
The differential operators $Q,\bar Q, D,\bar D$
defined by \eqref{Q_def} and \eqref{D_def}
 are kept unchanged, and $\theta$ and $\bar \theta$ do not depend on $t$. 
The supersymmetry transformation is defined by \eqref{super_transf}
replacing ${\cal F}(z)$ with ${\cal F}(z,t)$ where $\xi$ and $\bar \xi$ are $t$-independent parameters. 
The flowed vector and chiral superfields are defined by $V^\dag(z,t) = V(z,t)$ and  
the superchiral condition, $\bar D_{\dot\alpha} Q_{\pm}(z,t)=D_{\alpha} Q^\dag_{\pm}(z,t)=0$, respectively. 
The extended gauge transformation is defined in the same manner as \eqref{egauge} with flowed fields. 
Thus  all properties of superfields are inherited into the $t$-dependent superfields.
Hereafter the superfield actions \eqref{superfield_sym_action} and \eqref{superfield_mat_action}
are defined by replacing  $V(z) \rightarrow V(z,t)$ and $Q_\pm(z) \rightarrow Q_\pm(z,t)$.

In order to define a supersymmetric flow, we consider a norm \cite{Kikuchi:2014rla},
\begin{eqnarray}
\vert \vert \delta V \vert\vert^2 = \frac{1}{2} \int d^8 z \  {\rm tr} \left(
e^{-2V} \delta e^{2V} e^{-2V} \delta e^{2V} 
\right)(z)
\end{eqnarray}
which is invariant under $t$-independent super and extended gauge transformations. 
We can find a metric $g^{ab}(V)$:
\begin{eqnarray}
g^{ab}= 4\, {\rm tr}\, \left\{ T^a \left( \frac{{\cal L}_V^2}{\cosh(2{\cal L}_V)-1}\right)
T^b\right\} 
\end{eqnarray}
from $\vert \vert \delta V \vert\vert^2= \int d^8 z \, g_{ab}(V) \delta V^a \delta V^b$
and $g_{ac}g^{cb}=\delta_{a}{}^b$ where ${\cal L}_A B\equiv [A,B]$.

Thus, for the gauge multiplet, we define a supersymmetric gradient flow as 
\begin{eqnarray}
\partial_t {V}^a = -\frac{1}{2} g^{ab} \frac{\delta S_{\rm SQCD}}{\delta V^b}.  
\label{V_flow}
\end{eqnarray}
Since $g^{ab}$ is obtained from the invariant norm, 
the flow is covariant under 
$t$-independent supersymmetry and extended gauge transformations,  \eqref{super_transf} and  \eqref{egauge}.
We should notice that, without the metric, the gauge covariance is broken 
because the extended gauge transformation is nonlinear as in the case of general covariance. 
The supersymmetric covariance  holds without the metric because $g^{ab}$ is a function of superfield. .

For $Q_+$, a naive definition of supersymmetric flow would be 
$\partial_t Q_+  = \delta S_{\rm SQCD}/ \delta Q_+^\dag=  -\frac{1}{4} DD(e^{2V} Q_+)$. 
This is however 
incorrect as a supersymmetric flow for $Q_+$ because the RHS breaks the superchiral condition ($\bar D_{\dot\alpha} Q_+=0$)
and is not proportional to $\square Q_+$ in the free limit. 
For the definition of  $\delta/ \delta Q_+$ and detailed calculations, see
the derivation of field equations in Ref.~\cite{Wess:1992cp}.
Instead, if the RHS has an additional $\bar D^2$ factor,  
the superchiral condition is kept since $\bar D^3=0$ and $\square Q_+$ appears 
because $\bar D^2 D^2Q_+=16 \square Q_+$.

Therefore, for the matter multiplet, supersymmetric flows may be given by 
\begin{eqnarray}
&& \partial_t Q_+  
    = -\frac{1}{4} \bar D \bar D 
     \left(e^{-2V} \frac{\delta S_{\rm MAT}}{ \delta Q_+^\dag} \right),
\label{flow_Qp} \\
&& \partial_t Q_-   
  = -\frac{1}{4} \bar D \bar D 
    \left( \frac{\delta S_{\rm MAT}}{ \delta Q_-^\dag} e^{2V}\right).
\label{flow_Qm}
\end{eqnarray}
Note that $S_{\rm MAT}$ in \eqref{flow_Qp} and \eqref{flow_Qm} can be replaced by $S_{\rm SQCD}$
since $S_{\rm SYM}=S_{\rm SQCD}-S_{\rm MAT}$ does not have interactions with the matter multiplet.
These flow equations are covariant under 
$t$-independent supersymmetry and extended gauge transformations,  \eqref{super_transf} and  \eqref{egauge}.
We thus find the SQCD flow equations are given by \eqref{V_flow}, \eqref{flow_Qp}, \eqref{flow_Qm}
in the superfield formalism. 

For the superfield flow equations, the WZ gauge is broken at non-zero flow time because 
$\partial_t C = -D- (\phi^\dag_+ T^a \phi_+ - \phi_- T^a \phi_-^\dag) T^a \neq 0$ 
and $\partial_t \chi, \partial_tM ,\partial_t N$ are also non zero.
Therefore, we consider modified equations with an infinitesimal extended gauge transformation as 
\begin{eqnarray}
\label{V_flow_mod}
&&\partial_t {V}^a = - \frac{1}{2} g^{ab} \frac{\delta S_{\rm SQCD}}{\delta V^b} \ + \ \delta_\Lambda V^a
\end{eqnarray}
and
\begin{eqnarray}
\begin{split}
&
\partial_t Q_+  = -\frac{1}{4} \bar D \bar D 
\left(e^{-2V} \frac{\partial S_{SQCD}}{ \partial Q_+^\dag} \right) \ + \ \delta_\Lambda Q_+\\
& \partial_t Q_-  = -\frac{1}{4} \bar D \bar D 
\left( \frac{\partial S_{SQCD}}{ \partial Q_-^\dag} e^{2V}\right) \ + \ \delta_\Lambda Q_-
\end{split}
\label{Q_flow_mod}
\end{eqnarray}
where $\delta_\Lambda$ is an infinitesimal transformation derived from \eqref{egauge}.

We choose component fields of $\Lambda$ by hand such that
the WZ gauge maintains at non zero flow time: 
\begin{eqnarray}
\begin{split}
& A = \frac{D}{2} +\frac{1}{2}(\phi^\dag_+ T^a \phi_+ - \phi_- T^a \phi_-^\dag) T^a, 
\\
& \psi=  -\frac{1}{\sqrt{2}} \sigma^\mu D_\mu \bar\lambda + (\phi^\dag_+ T^a q_+ - q_- T^a \phi_-^\dag) T^a, 
\\
& F=   (\phi^\dag_+ T^a F_+ - F_- T^a \phi_-^\dag) T^a. 
\end{split}
\label{fixed_Lambda}
\end{eqnarray}
It is easily shown that  
$\partial_t C=\partial_t \chi =\partial_t M=\partial_t N=0$ 
from \eqref{fixed_Lambda} and the WZ gauge maintains at no zero flow time if it is set at $t=0$.

We finally obtain SQCD flow equations for the component fields as follows: 
For the gauge multiplet, we have
\begin{eqnarray}
\begin{split}
&\partial_t A_\mu = D^\rho F_{\rho \mu} 
-\lambda \sigma_\mu \bar\lambda - \bar\lambda \bar\sigma_\mu \lambda \\
&\hspace{1cm} 
+ i(\phi_+^\dag T^a D_\mu \phi_+ -D_\mu \phi_-T^a \phi_-^\dag - h.c. ) T^a  \\
&\hspace{1cm} 
+  (\bar q_+ \bar\sigma_\mu T^a q_+ +q_- \sigma_\mu T^a \bar q_- ) T^a, \\ \\
&\partial_t \lambda = -\sigma^\mu \bar\sigma^\nu D_\mu D_\nu \lambda
-[\lambda,D] \\
&\hspace{1cm} 
-\sqrt{2} \sigma^\mu ( \bar q_+ T^a D_\mu \phi_+  - D_\mu \phi_- T^a \bar q_- ) T^a \\
&\hspace{1cm} 
-\sqrt{2} i  ( F_+^\dag T^a q_+ -q_- T^a F_-^\dag ) T^a \\
&\hspace{1cm} 
- ( \phi_+^\dag \{\bar\lambda, T^a\} \phi_+ 
+ \phi_- \{ \lambda, T^a\} \phi_-^\dag) T^a, \\ \\
&\partial_t D = D^\mu D_\mu D 
+i ( D_\mu \lambda \sigma^\mu \bar\lambda 
- D_\mu\bar\lambda \bar\sigma^\mu \lambda - h.c.)  \\
&\hspace{1cm} 
-(\phi_+^\dag \{D,T^a\}\phi_+ + \phi_- \{D,T^a\} \phi_-^\dag )T^a \\
&\hspace{1cm}
+ 2 (D^\mu \phi_+^\dag T^a D_\mu \phi_+ -D^\mu \phi_-T^a D_\mu\phi_-^\dag ) T^a  \\
&\hspace{1cm} 
+ 2\sqrt{2} i (\bar q_+ T^a \bar\lambda \phi_+ + \phi_- \bar\lambda T^a \bar q_- 
-h.c. ) T^a \\
&\hspace{1cm} 
+ i (\bar q_+ T^a \bar\sigma^\mu D_\mu q_+
- q_- T^a \sigma^\mu D_\mu \phi_-^\dag - h.c. ) T^a  \\
&\hspace{1cm} 
-2 ( F_+^\dag T^a F_+ - F_- T^a F_-^\dag ) T^a.
\end{split}
\label{flow_gauge_1}
\end{eqnarray}
For the matter multiplet, we also have 
\begin{eqnarray}
\begin{split}
&
\partial_t \phi_+ = D^\mu D_\mu \phi_+ +  i\sqrt{2} \lambda q_+ \\
&\hspace{1cm} 
- (\phi_+^\dag T^a  \phi_+ - \phi_- T^a \phi_-^\dag) T^a \phi_+, \\ \\
&
\partial_t q_+ = -\sigma^\mu \bar\sigma^\nu D_\mu D_\nu q_+  \\
&\hspace{1cm} 
+i \sqrt{2} \lambda F_+ -\sqrt{2} \sigma^\mu \bar\lambda D_\mu \phi_+ -Dq_+ \\
&\hspace{1cm} 
-( \phi^\dag_+ T^a \phi_+  - \phi_- T^a \phi^\dag_- ) T^a q_+ \\
&\hspace{1cm} 
-2 ( \phi_+^\dag T^a q_+ - q_- T^a \phi_-^\dag ) T^a \phi_+, \\ \\
&
\partial_t F_+ = D^\mu D_\mu F_+ - 2 D F_+  \\
&\hspace{1cm} 
+\sqrt{2} (D_\mu\bar\lambda \bar\sigma^\mu q_+
- \bar\lambda \bar\sigma^\mu D_\mu q_+) -2 \bar\lambda \bar\lambda \phi_+ \\
&\hspace{1cm} 
-( \phi^\dag_+ T^a \phi_+  - \phi_- T^a \phi^\dag_- ) T^a F_+ \\
&\hspace{1cm} 
+2 ( \phi_+^\dag T^a q_+ - q_- T^a \phi_-^\dag ) T^a q_+\\
&\hspace{1cm} 
-2 ( \phi_+^\dag T^a F_+ - F_- T^a \phi_-^\dag ) T^a \phi_+,
\end{split}
\label{flow_matter_plus_1}
\end{eqnarray}
and 
\begin{eqnarray}
\begin{split}
&
\partial_t \phi_- = D^\mu D_\mu \phi_- -  i\sqrt{2} q_- \lambda \\
&\hspace{1cm} 
+  \phi_-T^a(\phi_+^\dag T^a  \phi_+ - \phi_- T^a \phi_-^\dag), \\ \\
&
\partial_t q_- = -\sigma^\mu \bar\sigma^\nu D_\mu D_\nu q_-  \\
&\hspace{1cm} 
-i \sqrt{2} F_- \lambda +\sqrt{2}  D_\mu \phi_- \sigma^\mu \bar\lambda +q_-D \\
&\hspace{1cm} 
+ q_- T^a( \phi^\dag_+ T^a \phi_+  - \phi_- T^a \phi^\dag_- ) \\
&\hspace{1cm} 
+2 \phi_- T^a( \phi_+^\dag T^a q_+ - q_- T^a \phi_-^\dag ) , \\ \\
&
\partial_t F_- = D^\mu D_\mu F_- + 2 F_- D  \\
&\hspace{1cm} 
-\sqrt{2} (D_\mu q_- \sigma^\mu \bar\lambda  
- q_- \sigma^\mu D_\mu \bar\lambda) -2 \phi_- \bar\lambda \bar\lambda \\
&\hspace{1cm} 
+ F_- T^a( \phi^\dag_+ T^a \phi_+  - \phi_- T^a \phi^\dag_- )  \\
&\hspace{1cm} 
-2 q_- T^a( \phi_+^\dag T^a q_+ - q_- T^a \phi_-^\dag ) \\
&\hspace{1cm} 
+2 \phi_- T^a( \phi_+^\dag T^a F_+ - F_- T^a \phi_-^\dag ).
\end{split}
\label{flow_matter_minus_1}
\end{eqnarray}
Note that these equations are given in the Minkowski space and, 
after the Wick rotation, 
they have correct sign for $\square \phi$ terms in the Euclidean space.

The equations \eqref{flow_gauge_1}, \eqref{flow_matter_plus_1} and  \eqref{flow_matter_minus_1} 
are gauge covariant for $t$-independent gauge transformations. 
We consider supersymmetry transformations for flowed fields 
extending \eqref{DF_super} naively to the flowed fields.
The straightforward calculations tells us that the time derivative and 
the supersymmetry transformation commute with each other 
up to a gauge transformation as
\begin{eqnarray}
&
[\partial_t,\delta_\xi]=\delta^g_{\omega},
\label{consistency_relation}
\end{eqnarray}
\begin{eqnarray}
\begin{split}
&
\omega \equiv  i\sqrt{2}(\bar\xi\bar\psi-\xi\psi)\\
& \ \   = i D^\mu (\xi \sigma_\mu \bar  \lambda +  \bar\xi \bar\sigma_\mu \lambda) \\
& \quad \ \  +i\sqrt{2}(\bar\xi \bar q_+ T^a \phi_+ + \xi q_- T^a \phi_-^\dagger)T^a\\
& \quad \ \  -i\sqrt{2}(\phi_-T^a\bar\xi\bar q_- + \phi_+^\dagger T^a\xi q_+)T^a,  
\end{split}
\end{eqnarray}
where $\delta^g_\omega$ is an infinitesimal gauge transformation given by 
an naive extension of \eqref{normal_gauge_transf} to the flowed fields.
Thus we find that the obtained flow is supersymmetric in the sense of \eqref{consistency_relation}
because the RHS of eq.~(\ref{consistency_relation}) vanishes for any gauge invariant observables.

Instead of \eqref{V_flow}, we consider another supersymmetric flow 
using the gradient of SYM action  instead of using the gradient of SQCD action
to define a gauge multiplet flow. 
The modified equation for the gauge multiplet is
\begin{eqnarray}
\label{V_flow_mod2}
&&\partial_t {V}^a = - \frac{1}{2} g^{ab} \frac{\delta S_{\rm SYM}}{\delta V^b} \ + \ \delta_{\Lambda^\prime} V^a
\end{eqnarray}
while ones for the matter multiplets are defined by \eqref{Q_flow_mod} with $\Lambda^\prime$. 
In this case, we choose the components of $\Lambda^\prime$ by hand as
\begin{eqnarray}
\begin{split}
& A = \frac{D}{2},
\\
& \psi=  -\frac{1}{\sqrt{2}} \sigma^\mu D_\mu \bar\lambda, 
\\
& F=  0,
\end{split}
\label{fixed_Lambda_prime}
\end{eqnarray}
to keep the WZ gauge. 
The change affects the matter-field flow through the $\delta_{\Lambda^\prime} Q_\pm$ term. 

A straightforward calculation tells us that
\begin{eqnarray}
\begin{split}
&\partial_t A_\mu = D^\rho F_{\rho \mu} 
-\lambda \sigma_\mu \bar\lambda - \bar\lambda \bar\sigma_\mu \lambda, \\
&\partial_t \lambda = -\sigma^\mu \bar\sigma^\nu D_\mu D_\nu \lambda
-[\lambda,D],\\
&\partial_t D = D^\mu D_\mu D 
+i ( D_\mu \lambda \sigma^\mu \bar\lambda 
- D_\mu\bar\lambda \bar\sigma^\mu \lambda - h.c.),
\end{split}
\end{eqnarray}
and 
\begin{eqnarray}
\begin{split}
&\partial_t \phi_+ = D^\mu D_\mu \phi_+ +  i\sqrt{2} \lambda q_+, \\
&\partial_t q_+ = -\sigma^\mu \bar\sigma^\nu D_\mu D_\nu q_+ 
+i \sqrt{2} \lambda F_+ \\
&\hspace{1cm} 
-\sqrt{2} \sigma^\mu \bar\lambda D_\mu \phi_+ -Dq_+, \\
&\partial_t F_+ = D^\mu D_\mu F_+ - 2 D F_+ \\
&\hspace{1cm} 
+\sqrt{2} (D_\mu\bar\lambda \bar\sigma^\mu q_+
- \bar\lambda \bar\sigma^\mu D_\mu q_+) -2 \bar\lambda \bar\lambda \phi_+, \\
&\partial_t \phi_- = D^\mu D_\mu \phi_- -  i\sqrt{2} q_- \lambda, \\
&\partial_t q_- = -\sigma^\mu \bar\sigma^\nu D_\mu D_\nu q_- 
-i \sqrt{2} F_- \lambda \\
&\hspace{1cm} 
+\sqrt{2} D_\mu \phi_- \sigma^\mu \bar\lambda +q_- D, \\
&\partial_t F_- = D^\mu D_\mu F_- + 2 F_- D  \\
&\hspace{1cm} 
-\sqrt{2} (D_\mu q_- \sigma^\mu \bar\lambda
- q_- \sigma^\mu D_\mu \bar\lambda) -2 \phi_- \bar\lambda \bar\lambda.
\end{split}
\end{eqnarray}
As in the case of eq. (\ref{consistency_relation}),
the consistency relation 
\begin{eqnarray}
&
[\partial_t,\delta_\xi]=\delta^g_{\omega^\prime},
\end{eqnarray}
\begin{eqnarray}
\begin{split}
&
\omega^\prime \equiv  i\sqrt{2}(\bar\xi\bar\psi-\xi\psi)
= i D^\mu (\xi \sigma_\mu \bar  \lambda +  \bar\xi \bar\sigma_\mu \lambda),
\end{split}
\label{consistency_relation_simple}
\end{eqnarray}
is also satisfied for these simplified equations.

\section{Summary}

We have derived supersymmetric gradient flow equations in  ${\cal N}=1$ SQCD. 
The flow equation is defined in both the superfields and the component fields of the Wess-Zumino gauge. 
The obtained flow is supersymmetric in the sense 
that the commutator of the flow time derivative and any supersymmetry transformation vanishes up to a gauge transformation.

In the YM flow, correlation functions of flowed gauge field are UV finite if the 4d YM theory is properly renormalized \cite{Luscher:2011bx}.
While, flowed fermion fields receive an extra wave function renormalization \cite{Luscher:2013cpa}. 
From a naive extension of these result,  
we expect that the flowed gauge multiplet shows the UV-finiteness
while the flowed matter multiplet may receive an extra renormalization factor.
The next step in our research is to show that this expectation is correct.
Then the Ferrara--Zumino multiplet 
on the lattice are constructed exactly on the lattice with the SQCD flow. 
These developments will be useful in lattice studies that verify the Seiberg duality in ${\cal N}=1$ SQCD. 

Supersymmetric gradient flow of ${\cal N}=2$ and ${\cal N}=4$ theories would be derived using techniques 
presented in this paper. 
If the ${\cal N}=1$ superfield formalism is used for ${\cal N}=2$ and ${\cal N}=4$ theories, 
the consistency relation \eqref{consistency_relation} may be broken for the whole of extended supersymmetry. 
In that case, how to keep the UV-finiteness will be unclear. 
Even so, the method given in this paper should help us understand extended cases.

\acknowledgement
{\it Acknowledgements}:
We would like to thank Nobuhito Maru and Mitsuyo Suzuki for their valuable comments.
This work is supported by JSPS KAKENHI Grant Numbers JP19K03853, JP20K03924, JP21K03537 and JP22H01222. 

%
\bibliographystyle{unsrt}
\bibliography{BibTex_template}

\begin{thebibliography}{10}

\bibitem{Luscher:2010iy}
Martin L{\"u}scher.
\newblock {Properties and uses of the Wilson flow in lattice QCD}.
\newblock {\em JHEP}, 08:071, 2010.
\newblock [Erratum: JHEP03,092(2014)].

\bibitem{Luscher:2011bx}
Martin L{\"u}scher and Peter Weisz.
\newblock {Perturbative analysis of the gradient flow in non-abelian gauge
  theories}.
\newblock {\em JHEP}, 02:051, 2011.

\bibitem{Luscher:2013cpa}
Martin L{\"u}scher.
\newblock {Chiral symmetry and the Yang--Mills gradient flow}.
\newblock {\em JHEP}, 04:123, 2013.

\bibitem{Suzuki:2013gza}
Hiroshi Suzuki.
\newblock {Energy-momentum tensor from the Yang-Mills gradient flow}.
\newblock {\em PTEP}, 2013:083B03, 2013.
\newblock [Erratum: PTEP2015,079201(2015)].

\bibitem{Makino:2014taa}
Hiroki Makino and Hiroshi Suzuki.
\newblock {Lattice energy-momentum tensor from the Yang-Mills gradient
  flow--inclusion of fermion fields}.
\newblock {\em PTEP}, 2014:063B02, 2014.
\newblock [Erratum: PTEP2015,079202(2015)].

\bibitem{Hieda:2017sqq}
Kenji Hieda, Aya Kasai, Hiroki Makino, and Hiroshi Suzuki.
\newblock {4D $\mathcal{N}=1$ SYM supercurrent in terms of the gradient flow}.
\newblock {\em PTEP}, 2017(6):063B03, 2017.

\bibitem{Kasai:2018koz}
Aya Kasai, Okuto Morikawa, and Hiroshi Suzuki.
\newblock {Gradient flow representation of the four-dimensional $\mathcal{N}=2$
  super Yang\textendash{}Mills supercurrent}.
\newblock {\em PTEP}, 2018(11):113B02, 2018.

\bibitem{Bergner:2019dim}
Georg Bergner, Camilo L\'opez, and Stefano Piemonte.
\newblock {Study of center and chiral symmetry realization in thermal
  $\mathcal{N}=1$ super Yang-Mills theory using the gradient flow}.
\newblock {\em Phys. Rev. D}, 100(7):074501, 2019.

\bibitem{Kikuchi:2014rla}
Kengo Kikuchi and Tetsuya Onogi.
\newblock {Generalized Gradient Flow Equation and Its Application to Super
  Yang-Mills Theory}.
\newblock {\em JHEP}, 11:094, 2014.

\bibitem{Kadoh:2018qwg}
Daisuke Kadoh and Naoya Ukita.
\newblock {Supersymmetric gradient flow in $\mathcal{N}=1$ SYM}.
\newblock {\em Eur. Phys. J. C}, 82(5):435, 2022.

\bibitem{Kadoh:2019glu}
Daisuke Kadoh, Kengo Kikuchi, and Naoya Ukita.
\newblock {Supersymmetric gradient flow in the Wess-Zumino model}.
\newblock {\em Phys. Rev. D}, 100(1):014501, 2019.

\bibitem{Wess:1992cp}
J.~Wess and J.~Bagger.
\newblock {\em {Supersymmetry and supergravity}}.
\newblock 1992.

\end{thebibliography}

\end{document}